\documentclass[conference]{IEEEtran}
\IEEEoverridecommandlockouts
\usepackage{cite}
\usepackage{amsmath,amssymb,amsfonts}
\usepackage{algorithmic}
\usepackage{graphicx}
\usepackage{textcomp}
\usepackage{amssymb}
\usepackage{xcolor}
\usepackage[utf8]{inputenc}
\usepackage[english]{babel}
\usepackage[utf8]{inputenc}
\usepackage[english]{babel}
\usepackage{amsthm}

\usepackage{multirow}
\usepackage{booktabs}

\begin{document}
\title{Deep Learning  with  Persistent Homology for Orbital Angular Momentum (OAM) Decoding}

\author{Soheil Rostami, Walid Saad, and Choong Seon Hong
\\ 
\thanks{S. Rostami is with Huawei Technologies Oy (Finland) Co. Ltd, Helsinki, Finland. E-mail: soheil.rostami1@huawei.com.}
\thanks{W. Saad is with  the   Bradley Department of Electrical and Computer Engineering, Virginia Tech, USA. E-mail: walids@vt.edu.}
\thanks{C. S. Hong is with Department of Computer Science and Engineering, Kyung Hee University, South Korea. E-mail: cshong@khu.ac.kr.}
\thanks{This research was supported by the U.S. National Science Foundation under Grant CNS-1909372.}\vspace{-55cm}}
\maketitle
\vspace{-55cm}
\begin{abstract}
Orbital angular momentum (OAM)-encoding has recently emerged as an effective approach for   increasing the channel capacity of free-space optical communications.  In this paper,    OAM-based  {decoding}  is    formulated as a supervised classification problem. To maintain lower    error rate in presence of severe    {atmospheric turbulence},   a new approach  that combines effective machine learning tools from persistent homology  and convolutional neural networks (CNNs) is proposed  {to decode the OAM modes}.   A Gaussian kernel with learnable parameters  is proposed in order to connect persistent homology to CNN, allowing the system   to extract and distinguish robust and unique topological features for the OAM modes. 
{Simulation results show that  the proposed approach achieves up to $20\%$ gains in classification accuracy rate over state-of-the-art of method based on only CNNs. These results essentially show  that  geometric and topological features play a pivotal  role in the OAM mode classification problem.} 
\end{abstract}


\section{Introduction}
Free-space optical (FSO) communication is an effective approach for fixed  point-to-point communication, such as backhaul connectivity and fiber backup   over distances up to several kilometers  \cite{Khalighi_survey}.  {In order to increase the transmission capacity of FSO communication, space division multiplexing can be exploited.  In particular, the use of the  orbital  angular momentum (OAM) of a light beam has been recently proposed to realize   space division multiplexing in FSO communication systems \cite{wang2012terabit}. {OAM is expected to   play a major role in many emerging communication systems  \cite{6G}.}
} 

Theoretically,  Laguerre-Gaussian (LG) beams can carry an infinite values of OAM modes,  by encoding a bit-tuple of information as superposition of  OAM modes \cite{wang2012terabit}.  {Therefore,   the   transmission capacity can be increased with orders-of-magnitude by allowing beams with different modes to be multiplexed together and transmitted over the same communication link.}   However, in  FSO communications,    the high sensitivity of the spatial structure of a light beam to atmospheric conditions such as turbulence can cause cross-talk among adjacent OAM modes. This, in turn, makes it challenging to perform error-free OAM mode detection particularly when a large   number of OAM modes  are used \cite{wang2012terabit}.
 
Conventionally, an optical solution based on coherent detection, known as  conjugate-mode sorting method   is applied for OAM detection as done in \cite{mair2001entanglement}. 
For the coherent detection of OAM modes, both   the  transmitter and receiver can use   spatial light modulators (SLMs).  In \cite{mair2001entanglement},   a first proof-of-concept experiment, based on coherent detection,  is developed to utilize OAM modes by  {defining an unlimited dimensional discrete   Hilbert space}. 
Recently,  neural network approaches have   been adopted in  \cite{Knutson}  and \cite{Doster:17} in order to enhance   OAM decoding by    relying only on an intensity image of the unique multiplexing patterns.  The authors in \cite{Knutson} proposed a deep neural network approach  capable of simultaneously differentiating $110$ OAM modes with   classification error rate of  less  than $30$\%. 
Meanwhile, the authors in \cite{Doster:17}, used convolutional neural networks (CNN) to differentiate $32$   OAM modes  with over $99$\% accuracy under high levels of turbulence. The work in \cite{Doster:17} also showed that this new method is robust to various environmental parameters.  However, the performance of the solutions proposed in \cite{Knutson}  and  \cite{Doster:17} degrades significantly for large  number of OAM modes and high levels of turbulence, thus motivating the need for new, turbulence-robust solutions. 

{Recently,  OAM communications has attracted significant attention  in the wireless communication literature. For  instance,  in \cite{cheng2018orbital}, the authors study the problem of enhancing spectrum efficiency for  multi-user access with different OAM modes  for  two-tier wireless networks. Meanwhile, in \cite{8712525}, the authors conducted important   experiments to characterize the OAM
phase properties for long-distance transmission.}

 The main contribution of this paper is a novel framework that exploits the powerful machine learning tools of \emph{persistent homology} \cite{zomorodian_2005} to enhance OAM mode detection, in presence of turbulence. In particular, the proposed approach combines a persistent homology-based input layer to a CNN-based OAM decoder. This  input layer allows inference of topological and geometrical information rather than the intensity images, for OAM mode detection.   Our results show   that exploiting topological features  is more effective    than phase fronts, applied in intensity images. In particular,  {simulation results show  that the proposed approach   outperforms CNN, in terms of   the classification accuracy rate,   by $10$\% in presence of severe atmospheric turbulence and a large number of OAM modes.}

 
\section{System Model}
\label{sec:system_model}
We consider an  OAM communication  system   composed of a single transmitter and receiver pair. 
The transmitter  communicates one out of $M$ possible $n$-bit length messages $s \in\mathcal{M}$, where $M=2^n$. To transmit each message,  $n$ different OAM modes $\{c_1,...,c_n\}$ are superpositioned\footnote{The throughput of the FSO communication link is linearly dependent on  $n$, and hence it is beneficial to increase $n$ as much as possible to achieve higher throughput for a given error rate.}, and the corresponding transmitted beam $\boldsymbol{x}\in \mathbb{C}^n $ is sent over the channel.  Hence, the transmitter can be seen as a  mapping function,  $\mathfrak{T}: \mathcal{M} \mapsto \mathbb{C}^n$. At the receiver,   the received beam $\boldsymbol{y}\in \mathbb{C}^n $ is  a noisy and  distorted version of the transmitted beam. The receiver must produce an estimate $\hat{s}$ of the original message $s$,   and, hence, the receiver can be seen as  a mapping function,  $\mathfrak{R}:\mathbb{C}^n \mapsto \mathcal{M}$. Both $\boldsymbol{x}$ and $\boldsymbol{y}$ are vectors  with each element   corresponding, respectively, to the intensity level of the transmitted or received beam's  pixel. The length of $\boldsymbol{x}$ and $\boldsymbol{y}$ equals to the number of pixels per beam $p$,  which  depends on the pixel size  of a charge-coupled device   {image sensor}.

The considered OAM communication system must reduce the message  error rate ($P_e=\text{Pr}  {(\hat{s}\neq s)}$)   {for a  high  number of OAM modes} and severe turbulence levels.  {Moreover, some   OAM modes are more robust to   atmospheric turbulence, and, thus, the $n$ OAM modes that are used for
 encoding a message can be selected in such a way to have the least sensitivity to the channel conditions.}  Here, we assume that the set of OAM modes, i.e. $\{c_1,...,c_n\}$ is fixed.  

{A CNN learning technique  can be used  for OAM decoding with the objective of minimizing the message error rate.  Such a neural network can be modeled  as a function composition  chain of  functions, $f(\boldsymbol{y}) = {a}_L\circ {q}_L \circ {a}_{L-1} \circ {q}_{L-1} \circ \dots\circ {a}_{1} \circ {q}_1(\boldsymbol{y})$; where ${a}_j$ and ${q}_j$ are   convolutional (conv1,...,convL) and pooling (pool1,...,poolL) functions, respectively; and 
$L$ is the number of layers in the network. 
The convolution filters apply convolution operations to the input, and then   pooling applies a moving window   to choose the maximum value over a local region.
The final layer (based on softmax  activation) of the network represents the $M$ messages. Once trained, during the testing 
phase, the CNN   yields a set of  probabilities $\{f_{1},...,f_{m},...,f_{M}\}$, where  $f_{m}$ is essentially the probability that the CNN's input message  is classified as  message $m$.} 
Message $i$ is detected as the received message if {$f_{m}<f_{i}$ for all  $m\in \mathcal{M}$ and $m\neq i$.} 

Although CNN can be effective for improving OAM reception as shown in \cite{Doster:17}, relying solely on CNN for the  purpose of OAM  detection can fail if the number of OAM modes is large. In particular, CNN   performs its decoding by relying solely on  the intensity profiles while ignoring the topological structure of each OAM mode. However, in practice, each OAM mode, also known as a topological charge, possesses a unique  twisted structure. Such topological structures contain more discriminate  features than intensity profiles. Hence, in order to enhance the effectiveness of OAM decoding, it is desirable to design new learning techniques that can exploit such topological structures, which has not been done in prior works  \cite{Knutson, Doster:17}. In this context, persistent homology  has recently emerged as a powerful tool for addressing the problem of topological  feature detection and shape recognition.  By means of  persistent homology, a shape is represented  with a family of simplicial complexes, indexed by a proximity parameter  \cite{zomorodian_2005}. This converts different shapes  into global topological objects. The output of persistent homology is  in the form of a parameterized version of a Betti number, and can be illustrated by a persistence  diagram  \cite{zomorodian_2005}.  Depending on the   combination of OAM modes used in the transmitted beam,   the received beam will exhibit a unique topological structure which differentiates it from other feasible combinations of OAM modes. Therefore,   persistent homology can be used to extract such features which can, in turn, be exploited by the CNN to classify the received message. {Table \ref{tab:vari} summarizes our notations.}
 
 \begin{table}[!t] 
 \centering
\renewcommand{\arraystretch}{1.2}
\caption{{Summary of main notations.}} 
\label{tab:vari}
\begin{tabular}{ccccc}
\cline{1-2}
\multicolumn{1}{|c|} {\begin{tabular}[c]{@{}l@{}}\textbf{Variable}\end{tabular}} & \multicolumn{1}{c|}{\textbf{Definition} }                                                              &  &  &  \\ \cline{1-2} 
\multicolumn{1}{|c|}{ $\boldsymbol{\mu_i}=[\mu_{i,1},\mu_{i,2}]^T$ } & \multicolumn{1}{c|}{{Location parameters  of Gaussian Kernel}}                                                               &  &  &  \\ \cline{1-2}
\multicolumn{1}{|c|}{ $\sigma_i$} & \multicolumn{1}{c|}{{Standard deviation of Gaussian Kernel}}                                                               &  &  &  \\ \cline{1-2}
\multicolumn{1}{|c|}{ $\nu$ } & \multicolumn{1}{c|}{{\begin{tabular}[c]{@{}c@{}}Parameter  to control the effect of the\\ persistence and computational complexity\end{tabular}}}         
      &  &  &  \\ \cline{1-2}
      \multicolumn{1}{|c|}{ $\mathcal{B}$ } & \multicolumn{1}{c|}{
      {\begin{tabular}[c]{@{}c@{}}Persistence diagram (can be \\ visualized as  multisets of intervals) \end{tabular}}}         
      &  &  &  \\ \cline{1-2}            \multicolumn{1}{|c|}{ $|.|$ } & \multicolumn{1}{c|}{{Cardinality of a set}}         
      &  &  &  \\ \cline{1-2}
\multicolumn{1}{l}{}    & \multicolumn{1}{l}{}             &  &  & 
\end{tabular}
\vspace{-8 mm}
\end{table}

Due to space limitation, the mathematical intricacies of persistent homology are not provided in detail; we refer the reader to  \cite{zomorodian_2005}  for  algebraic-topological aspects  of persistent homology. Next, we propose a novel framework that employs persistent homology as an input layer for CNN   in order to probe topological features  of OAM modes and enhance the OAM receiver performance.

\section{Proposed Framework}
\label{sec:Solution}
The utilization of unique topological structures of OAM modes is promising to enhance OAM mode detection  {because the  {topological features} are independent of any particular coordinate. 
As result, no optical alignment process   is needed.}  Furthermore, the  topological features of OAM beams are deformation invariant, i.e. their topology does not change if the shape is stretched or compressed, due to  atmospheric turbulence. Also, topological analysis of OAM beams   enhances robustness  to noise, enabling   long distance communication. 

Due to the superposition of $n$-orthogonal topological charges,  each transmitted beam $\boldsymbol{x}$ and hence corresponding received beam $\boldsymbol{y}$ will exhibit  a unique topological structure.  {In this regard,  persistent homology will allow us to replace each point of an OAM topological structure with a solid sphere and, then, record the topological features of the union of these solid spheres as a function of their radius.} When  the radius increases, new components can be created, or existing components can merge together, or can be eliminated. 
The $0^{th}$ order topological feature captures the connected components, the $1^{st}$ order topological feature captures regions forming a loop structure, and $2^{nd}$ order topological feature captures regions forming a void structure.  {The radius at which the $j^{th}$ component, referred to   as $p_j$, is created is called its \emph{birth time} $b_j$ and the radius at which   component $p_j$  is eliminated, or merged with another component  that has an earlier birth time, is called its \emph{death time} $d_j$}.  {For our OAM system, persistent homology detects the  connected components ($0^{th}$ order), loops ($1^{st}$ order), and voids ($2^{nd}$ order) of the received beam. Meanwhile,  the corresponding $(b_j,d_j)$-tuples  measure  how persistent  the received beam's topological features are. In other words, the difference $(d_j - b_j)$    is   known as the \emph{lifetime} of the topological feature and measures its robustness and prominence.}

\begin{figure}[t]
\centering
\includegraphics[scale=.8]{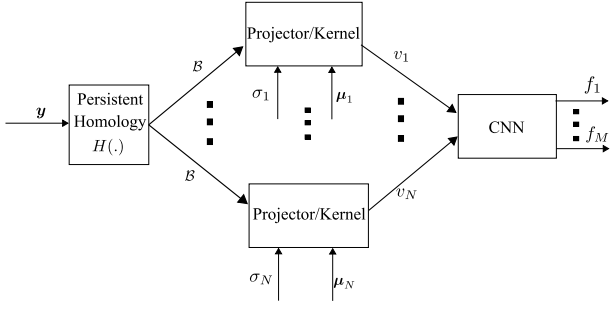}
\caption{The overall functionality of the proposed method used by an OAM receiver to classify OAM modes, both  projector and CNN are  trained as a single neural network. }
\label{fig:overall}\vspace{-7 mm}
\end{figure}

 {The persistence diagram  
 can be visualized as  multisets of intervals   $\mathcal{B} =\{ {p}_j=(b_j,d_j):0<b_j<d_j,j=1,...,|\mathcal{B}| \} $.  We define a \emph{persistent homology}   as a mapping function ${\mathfrak{H}}$, which maps the received beam  $\boldsymbol{y}$ to its corresponding persistence diagram $\mathcal{B}$, i.e. $\boldsymbol{y}\overset{\mathfrak{H}}{\mapsto}   \mathcal{B}$.} 
The  persistence diagram  $\mathcal{B}$  does not possess a Hilbert space structure due to its  unusual structure as   multisets. As result,   machine learning algorithms (e.g. CNN) operating on a Hilbert space cannot be applied  to persistence diagrams.  However, a persistence diagram  can embed the set of persistence diagrams into a Hilbert space   by applying   \textit{kernels} to map the topological features into machine learning compatible representation \cite{Kusano}.   
However, such   kernels are pre-defined and fixed, and therefore agnostic to the learning stage of the supervised classification.  For our case, we propose  a modified version of a Gaussian kernel with learnable parameters. In particular, the parameters of our kernel are  tuned   using back-propagation. 

For the CNN of our OAM decoder, we introduce a persistent homology-based input layer that is defined  by    projecting any  {component} ${p}_j$ with respect to a collection of kernels. Therefore,  each   component ${p}_j$, 
is transformed to a single value ${z}_j$ (i.e. ${p}_j\overset{G_{\sigma_i,\boldsymbol\mu_i}}{\mapsto} {z}_j$) as follows,
 $G_{\sigma_i,\boldsymbol\mu_i}({p}_j)=e^{-\frac{\lVert p_j-\boldsymbol{\mu_i} \rVert}{2{{\sigma_i}^2}}}$, where
  $\boldsymbol{\mu_i}=[\mu_{i,1},\mu_{i,2}]^T$ is a location parameter and $\sigma_i>0$ is standard deviation. Both are learned using a backpropagation learning method during the training stage.

Due to fact that components close to the diagonal of persistence diagram are mainly noise,
we can remove these components in our analysis,  i.e. $z_j=0$ if $d_j-b_j< \nu$ where $\nu$ is a constant. Consequently,  not only the contribution of noise is discounted but also {computational complexity is reduced.}

 The  persistence diagram $\mathcal{B}$ per kernel is projected into a single value  $v_i=\sum_{j=1}^{|\mathcal{B}|} G_{\sigma_i,\boldsymbol\mu_i}({p}_j)$.
\noindent By  projecting  $\mathcal{B}$ with respect to  $N$ kernels with different parameters, the   {\emph{topological feature vector} $\boldsymbol{v}$ of the received OAM beam $\boldsymbol{y}$ can be constructed, i.e. ${z}_j\overset{\mathfrak{v}}{\mapsto} \boldsymbol{v}$, where $\boldsymbol{v}=[v_1,...,v_{N}]^T$.}

The end-to-end operation of  the  framework is     shown in Fig. \ref{fig:overall}. First,    $\mathcal{B}$ is computed using any standard   persistent homology algorithm (e.g. \cite{zomorodian_2005}) over the received OAM beam $\boldsymbol{y}$, and then $\mathcal{B}$ is projected to $\boldsymbol{v}$ which itself is an input layer for the CNN. Therefore, the overall function of the resulting neural network can be written as follows, 
\begin{equation}
f(\boldsymbol{y}) =  {a}_L\circ  {q}_L \circ  {a}_{L-1} \circ  {q}_{L-1} \circ \dots\circ  {a}_{1} \circ  {q}_1\circ  \mathfrak{v}  \circ  \mathfrak{H} (\boldsymbol{y}),
\end{equation}
\noindent {where   ${\sigma_i}$ and $\boldsymbol{\mu}_i$ (for all $i\in\{1,...,N\}$) and corresponding parameters of $L$ layers are learned during training.} 

\begin{figure}
\centering
  \includegraphics[width=0.37 \textwidth]{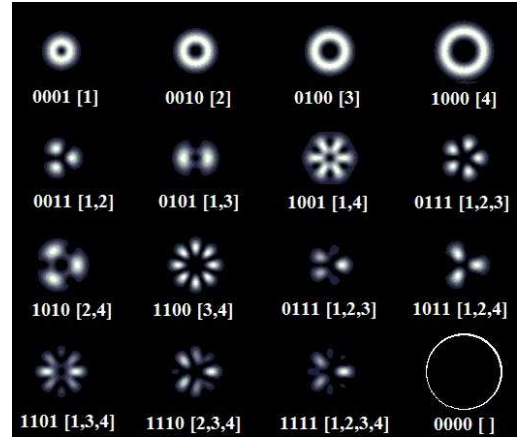}
  \caption{{The transmitted OAM beam when $M=16$. }}
  \label{fig:figo}\vspace{-4 mm}
\end{figure}

{An example of a computer-generated   phase diagram
  without turbulence for superposition of an OAM mode set  $\{1,2,3,4\}$  is shown in Fig. \ref{fig:figo}. The caption of each sub-image is the encoded 4 bit-length  message  and
the modes that are active (set of integers in brackets). For instance,  $0101 [1,3]$ means that for four-bit message $0101$, OAM modes $1$ and $3$ are superpositioned and transmitted.}
Now, based on  Fig. \ref{fig:figo}, we can illustrate the reasons why the proposed approach, that combines persistent homology and CNN, is more effective  than CNN alone,   for   OAM communications. In this regard, the received beam of OAM communication has a unique “twisted”  structure  
 (e.g. see Fig. \ref{fig:figo}). The persistent homology acts as a topological feature selector which can automatically   choose those topological OAM features   that contribute  the most to the output.  However, without   persistent homology, the trained CNN model will rely on the irrelevant features and noise, which can then reduce the    accuracy of the OAM decoder.
Furthermore,  by using persistent homology, each received OAM beam can be properly categorized to its unique topological structures through topological feature vector $\boldsymbol{v}$. Last but not least, the proposed method reduces training time and algorithm complexity by removing the irrelevant features and the components that most likely are linked to noise.

\begin{figure}[!t]
\centering
  \includegraphics[width=0.455 \textwidth]{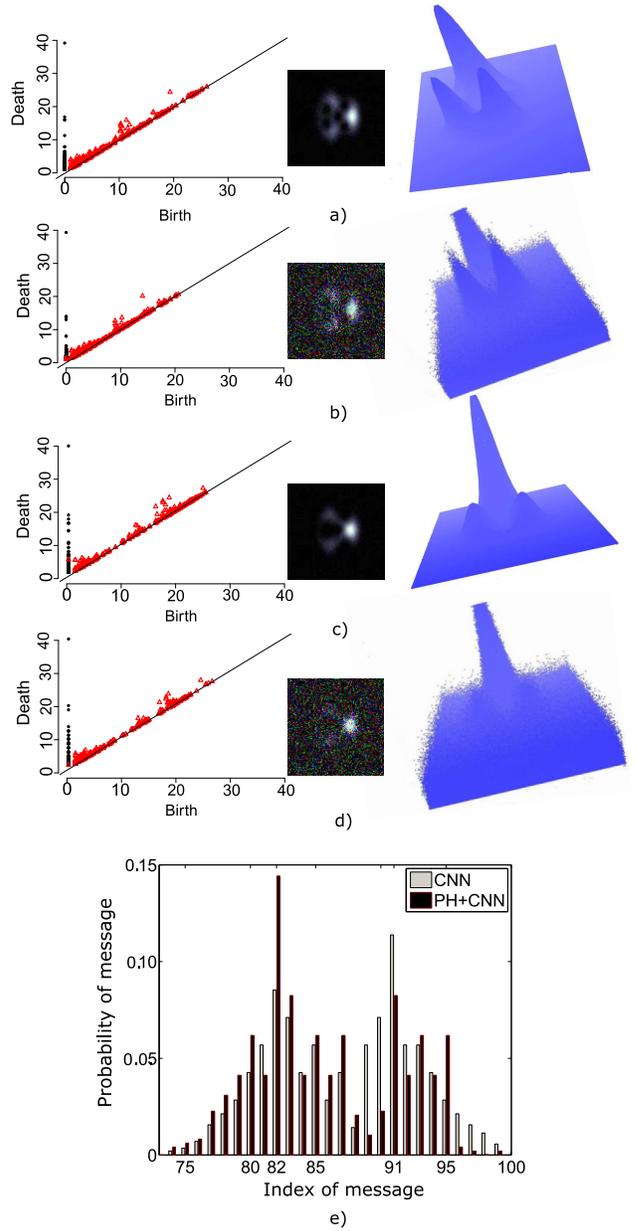}
  \caption{ The OAM decoding of $M=128$; due to importance,  the $1^{st}$ and  $2^{nd}$  order topological feature are shown with black   and red, respectively;  a)   the $82^{nd}$ message is shown at the transmit side ($\boldsymbol{x}$); b)   the $82^{ed}$ message is shown at the receive side ($\boldsymbol{y}$); c)  the $91^{st}$ message is shown at the transmit side ($\boldsymbol{x}$); d)  the  $91^{st}$ message is shown at the receive side ($\boldsymbol{y}$); e) $f_m$ is illustrated for both proposed method PH+CNN and CNN. }
  \label{fig:fig11}\vspace{-5 mm}
\end{figure}

In order to intuitively explain how the proposed method   outperforms conventional CNN, the persistence diagram, phase front and topological structure of a pair of $8$-bit length OAM messages both in transmit and receive are provided in Fig. \ref{fig:fig11}. In particular, two OAM beams corresponding to the $82^{nd}$ and $91^{st}$ messages are chosen. The main reason for choosing  the pair as example  is that, CNN frequently misclassifies $82^{nd}$ message with $91^{st}$. In contrast, our proposed method  classifies this particular example correctly most of time. It is clear   that for both received messages,   their persistence diagrams, especially the $2^{nd}$  order topological features (shown in black)  do not vary much compared to the phase diagrams. As result, the CNN  which relies on   phase diagrams,  will misclassify the received OAM beam.  Significant features of 3D topological shape (shown in blue) are far from the diagonal in the persistence diagram, {and their corresponding points in the diagram do not change significantly. Furthermore, Fig. \ref{fig:fig11} e) illustrates the output layer (the probabilistic values between $0$ and $1$) resulting from  the proposed framework and   conventional CNN, when the transmit message's index is $82$. Clearly, in the proposed framework, $f_{82}$   is much higher than other  $f_m$  for all $1 \leq m \leq 128$ and $m\neq 82$. However, in case of conventional CNN,  $f_{91}$ is   higher than  other $f_m$  for all $1 \leq m \leq 128$ and $m\neq 91$, and thus the CNN  misclassifies the $82^{nd}$ message as the $91^{st}$.}
 
{Our 
network is based on  Alexnet \cite{alexa} and is composed of 5 conv-pooling  layers (explained in Section \ref{sec:system_model})
and 3 fully connected (fc) layers. The detailed network architectures of the proposed method (PH, conv1, pool1,..., conv5, pool5, fc1, fc2, fc3) and baseline (conv1, pool1,..., conv5, pool5, fc1, fc2, fc3)  are shown in Table \ref{table:nonlin}. In order to remove the dependency of the algorithm's running time  on hardware,  the floating-point instruction (FLOP) counts  are calculated. The FLOP
counts of  backward propagation layers  are listed in Table \ref{table:nonlin}, from which we can see that processing overhead of computing persistent homology (PH) is much lower than the rest  of the convolutional layers. Also, Table \ref{table:rtr} shows that the running time and FLOP  counts of the proposed method (including persistent homology plus Alexnet-based CNN) are just 0.3\% and 0.5\%  higher than the baseline (Alexnet) when both are running on GTX1080, respectively. }
\begin{table}[!t]
\scriptsize
\caption{{Parameter analysis of layers used in the proposed method (PH+CNN)  and baseline (CNN: conv1, pool1,..., conv5, pool5, fc1, fc2, fc3)  when $N=1000$.}} 
\centering 
\begin{tabular}{c|c|c c c   c  c} 
\hline\hline 
Proposed method &Baseline &Layer & Kernel &   Parameters & FLOPs  \\ [0.5ex] 
\hline 
PH&- &PH & $1\times 1$ &  3 K    &  1 G   \\
\hline
& &conv1 & $11\times 11$  & 35 K &  40 G   \\
& &pool1 & $3\times 3$    &   0  &  1 G  \\
& &conv2 &  $5\times 5$   &  615 K     & 120 G    \\
& &pool2 &  $3\times 3$   &   0    &   1 G\\
CNN& CNN&conv3 &  $3\times 3$   &   885 K     & 40 G  \\
& &conv4 &   $3\times 3$  &    1.33 M   &  60 G   \\
& &conv5 &  $3\times 3$   &   885 K   &  60 G  \\
& &pool5 & $3\times 3$    &   0   &   1 G  \\
& &fc6   & $1\times 1$    &   38 M     &  10 G   \\
& &fc7   & $1\times 1$    &   17 M    &  5 G   \\
& &fc8   & $1\times 1$    &   4 M    &   2 G  \\

\hline 
\end{tabular}
\label{table:nonlin}\vspace{-5 mm} 
\end{table}

\begin{table}[!t]
\scriptsize
\caption{{Overall FLOP counts, number of parameter and running time    of proposed method (PH+CNN) and baseline (CNN)  when $N=1000$.}} 
\centering 
\begin{tabular}{c c   c  c c} 
\hline\hline 
Method   &   Parameters & FLOPs & Running Time \\ [0.5ex] 
\hline 
PH+CNN   &  62.753 M    &  341 G  & 42.1 ms\\
CNN   & 62.75 M  &  340 G& 42 ms\\

\hline 
\end{tabular}\vspace{-3 mm}
\label{table:rtr} 
\end{table}

\section{Simulation results and Analysis}
\label{ref:sim}
 In this section, we  demonstrate  the performance of the proposed method in terms of accuracy  {($\text{Pr}  {(\hat{s}= s)}=1-P_e$)},  {and we compare it with   the   CNN-based method (Alexnet) of  \cite{Doster:17}.} For the purpose of training and testing the framework,   the first-$n$-adjacent OAM modes, 
 ($n \in\{6,...,16\}$)   are  generated and superpositioned, numerically.
 
 Additionally,  the effect of varying the turbulence level and the message length (which is equal to the number of OAM modes) on the accuracy rate $(1-P_e)$  are investigated. Turbulence causes a random phase  along the propagation path of the beam    \cite{Andrews}, and the turbulence level, denoted by $T$, can be quantified  as the ratio between  linear dimension of the SLM and the Fried’s parameter  \cite{Andrews}. 
 We compare the performance of the proposed method, called “PH+CNN”, with
those achieved by the CNN in \cite{Doster:17}, and we call it “CNN”.


{Our dataset is generated numerically by using the   “basic paraxial optics toolkit” in MATLAB \cite{bworld}, and it 
consists of a balanced group size (per OAM) of $10000$ phase diagrams that contains  a variable amount  of turbulence.  
To train the CNN, we split the
data collected into two separate sets a training set ($85\%$) and testing set ($15\%$) of the overall dataset. The training and testing sets are completely independent of one another and do not share any
turbulence realizations. The training and testing set for the PH+CNN  use the same turbulence realizations as those used for the CNN.} {In our simulations, the most meaningful results were found by empirically setting $\nu$ to $0.1$. Therefore, all results are reported for $\nu=0.1$. By adjusting $\nu$, we can control and discount the effect of the points with low persistence and computational complexity (explained in Section  \ref{sec:Solution}).}  Furthermore,  for  computing persistent homology, the  “R package TDA” \cite{Fasy2014IntroductionTT} is used.

\begin{figure}
\centering
  \includegraphics[width=0.36\textwidth]{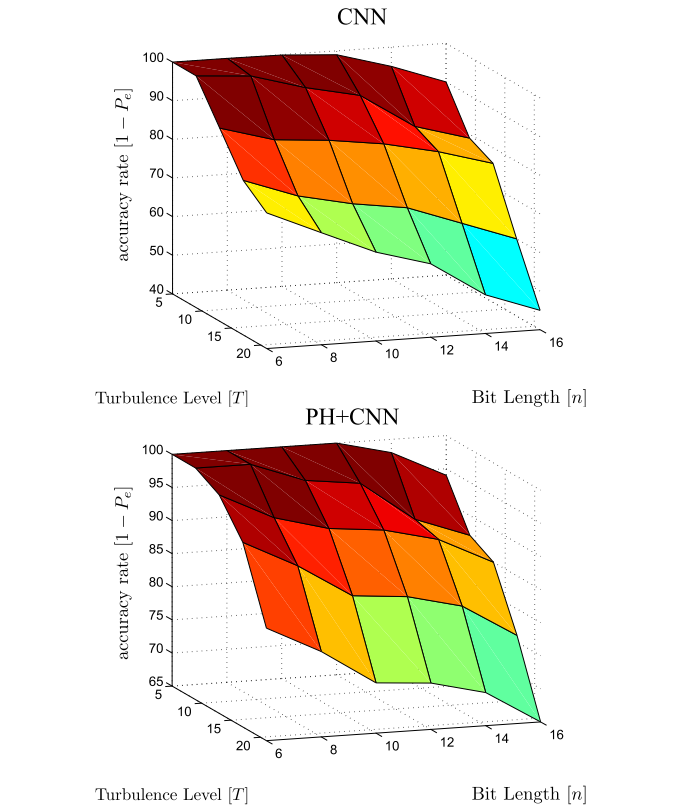}
  \caption{{Decoding accuracy of    CNN (top) and   PH+CNN (bottom) as a function of  bit lengths and turbulence levels.}}
  \label{fig:fig2}\vspace{-3 mm}
\end{figure}

{Fig.  \ref{fig:fig2}   shows the accuracy of decoding   for different turbulence levels and bit lengths of CNN and PH+CNN.   Fig.  \ref{fig:fig2} shows that,  PH+CNN outperforms   CNN  for different turbulence levels and bit lengths.       Fig.  \ref{fig:fig2} shows that  the performance  of both methods degrades rapidly with increasing   turbulence level and bit length.     However,  for higher turbulence levels, the performance of PH+CNN is much better than CNN while  they perform similarly    at     low turbulence levels. It can be seen that for a severe turbulence level of $21$,   for a message with $16$-bit length, PH+CNN yields up to $20$\% better accuracy than CNN.     This demonstrates that the PH+CNN method can allow a larger message size or OAM number and can increase the capacity further even for high turbulence levels.} 


\section{Conclusions}
\label{conc}
In this paper,  we have proposed a new approach that combines  persistent homology and CNN to decode OAM modes. We have shown that, after training the proposed method on set of     each superpositioned OAM mode, a high accuracy, rate even with large number of  OAM modes,   in presence of severe    {atmospheric turbulence} can be achieved. Numerical results have shown a substantial increase in the number of simultaneously-discriminated OAM modes  under turbulence channel     with better accuracy than CNN. Future work can consider the design of an autoencoder to select   OAM modes with highest discriminative topological features as the messages, for   given channel conditions, data rates and error rate requirements, adaptively. 

\bibliographystyle{IEEEtran}
\bibliography{IEEEabrv,Master}
\end{document}